\documentclass[twocolumn,nofootinbib,amsmath,amssymb,aps,prd,balancelastpage,superscriptaddress]{revtex4-1}

\usepackage{color}
\usepackage[active]{srcltx}
\usepackage{amsmath,amsfonts,amssymb,amsthm,amstext,amscd,eucal,srcltx}
\usepackage{epsfig,graphicx,bm}
\usepackage{epstopdf, epsf}
\usepackage{dcolumn}
\usepackage{hyperref}

\newcommand{\be}{\begin{equation}}
\newcommand{\ee}{\end{equation}}

\newcommand{\bse}{\begin{subequations}}
\newcommand{\ese}{\end{subequations}}
\newcommand{\bea}{\begin{eqnarray}}
\newcommand{\eea}{\end{eqnarray}}
\newcommand{\ba}{\begin{array}}
\newcommand{\ea}{\end{array}}
\newcommand{\bc}{\begin{center}}
\newcommand{\ec}{\end{center}}
\newcommand{\vect}{\mathbf}

\begin{document}
\vspace*{3mm}

\title{Solar System and Atomic Stronger Bounds on Exotic Dyonic Matter and Non-associative Quantum Mechanics}

\author{Andrea Addazi}
\email{andrea.addazi@lngs.infn.it}
\affiliation{Center for Field Theory and Particle Physics \& Department of Physics, Fudan University, 200433 Shanghai, China}

\author{Antonino Marcian\`o}
\email{marciano@fudan.edu.cn}
\affiliation{Center for Field Theory and Particle Physics \& Department of Physics, Fudan University, 200433 Shanghai, China}

\begin{abstract}
\vspace*{6mm}
\noindent
We derive stronger bounds on the magnetic monopole charge of the proton, and hence on non-associative quantum mechanics from measurements of the magnetic fields of the Earth, the Moon and Mars. Limits are several orders of magnitude stronger than the ones inferred from the Hydrogen atomic levels, assuming that electrons do not retain a magnetic monopole charge. Conversely, we show how to estimate bounds on the magnetic charge of a proton correctly, when the magnetic monopole charge of the electron is taken into account. 
\end{abstract}

\maketitle

{\it Introduction.--} Recently, Bojowald {\it et al.} have shown that strong experimental bounds on the magnetic monopole charge of the proton in non-associative quantum mechanics may be attained from measurements of the levels of Hydrogen-like atoms \cite{Bojowald:2018qqa}. Here we scrutinize and elaborate on the arguments of Ref.~\cite{Bojowald:2018qqa}, and provide stronger bounds on dyonic\footnote{Dyons are particles provided with both electric and magnetic monopole charges.} matter. As can be seen in Ref.~\cite{Bojowald:2018qqa}, this percolates in setting constraints on non-associative quantum mechanics.

\vspace{2mm}

{\it Magnetic monopoles and non-associative quantum mechanics.--} We start from the standard formulation of non-associative quantum mechanics, encoded in the commutator 
\begin{equation}
\label{ppij}
[p_{j},p_{k}]=\imath e \hbar \sum_{l=1}^{3}\epsilon_{jkl}B^{l}, 
\end{equation}
where we denote coordinates with $x_i$ and conjugated momenta with $p_i$, and the magnetic field as $B^i$. We leave the standard commutator $[x_{j},p_{k}]=\imath\hbar \delta_{jk}$ unchanged. The charge of the electron, $e$, appears in its absolute value. This construction may arise within the case of the magnetic field of a Dirac monopole. Let us assume that the proton has a monopole magnetic charge, and focus on hydrogen-like atoms. We may proceed considering the two cases: i) the proton only retains a magnetic monopole charge; ii) both the proton and the electron possess a magnetic charge, equal and opposite in sign.   

As an immediate consequence of the spherical symmetry and the Gauss theorem, the magnetostatic field of the monopole source has the form
\begin{equation}
\label{Bg}
\vect{B}=g(\vect{r})\vect{r}\, , 
\end{equation}
where 
\begin{equation}
\label{dvectr}
g(\vect{r})=Q_{p}(\vect{r})/4\pi r^{3}\,,
\end{equation}
a magnetic charge being enclosed into the proton radius as 
\begin{equation}
\label{Qmm}
Q_{p}(\vect{r})=4\pi \int_{0}^{R_{\rm Proton}} \nabla \cdot \vect{B}(r) r^{2}dr\, . 
\end{equation}
For simplicity, along Ref.~\cite{Bojowald:2018qqa}, we may assume the monopole to be point-like, {\it i.e.} $Q(r)=g={\rm const}$.  
%
The framework considered by Bojowald {\it et al.} falls in the first class (non-dyonic electrons). It corresponds to a particular case studied by Malkus in Ref.~\cite{M}, and amounts to a spin-less point-like interaction of the electron with the magnetic monopole of the proton. The interaction of the electron with the proton was considered in its full generality in Ref.~\cite{M}, accounting also for the interaction between the magnetic momentum of the electron and the magnetostatic field generated by the proton. 

\vspace{2mm}
 
{\it Tighter experimental constraints for case i).--}
We immediately notice that assigning a magnetic monopole charge only to the proton, and not to the electron, allows us to derive much stronger (by several orders of magnitude) constraints than the limits discussed in Ref.~\cite{Bojowald:2018qqa}. For this case, as estimated in Ref.~\cite{PT}, the proton magnetic charge would be already constrained by the bound $g\leq 10^{-43}$ (A m system), corresponding to $g\leq 3 \times 10^{-35}g_{\rm Dirac}$. We immediately provide novel constraints when considering other planets of our solar system. From the characteristics of Mars (with polar radius $R\!\sim \!0.533 \, R_{\odot}$, mass $M\!\sim\! 0.107\, M_{\odot}$ and average magnetic field $B\!\sim\! 10 \, B_{\odot}$ ), the limit on the magnetic monopole charge is found to be weaker by a factor $5.15$ with respect to the bounds provided in Ref.~\cite{PT}. But from the Moon (with $R\!\sim \!0.273 \, R_{\odot}$, $M\!\sim\! 0.0123\, M_{\odot}$ and $B\! \lesssim \! 10^{-2} \, B_{\odot}$) a stronger value by a factor $0.246$ can be recovered. 

\begin{figure}[ht]
\centerline{ \includegraphics [width=1.15\columnwidth]{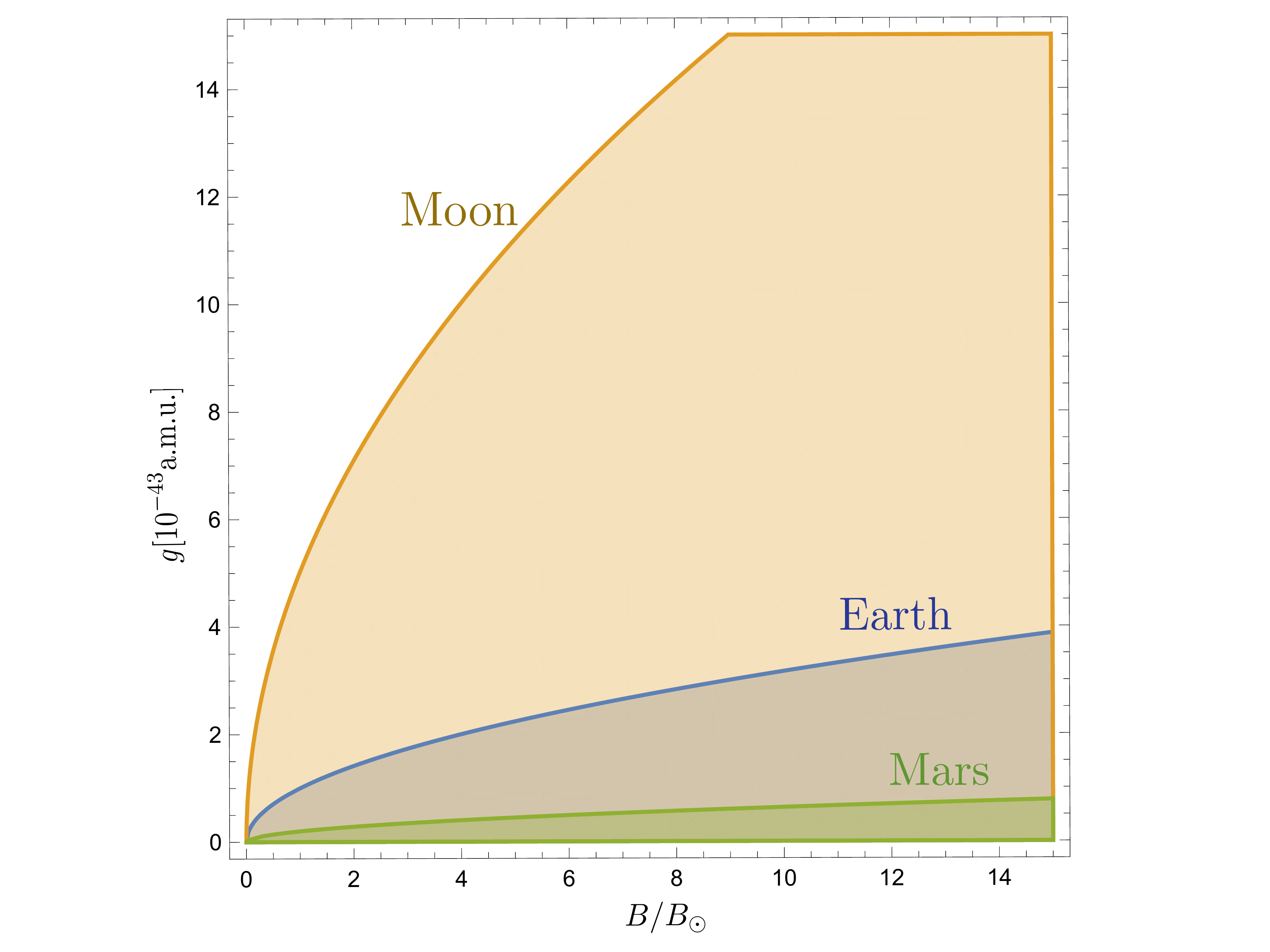}}
\caption{We display region exclusion plots for the limits on the magnetic monopole charge, as provided by the Moon (brown line) the Earth (blue line) and Mars (green line). }
\end{figure}

Within a different framework than the one contemplated in Ref.~\cite{Bojowald:2018qqa}, one should then consider, based on trivial stability arguments, that the magnetic monopole charge of the electron must be opposite in sign and equal in absolute value to the magnetic charge of the proton. The relative contribution that would arise from the magnetostatic interaction would be then of the same order of magnitude of the interaction related to the non-associativity of the momenta, arising from Eq.~\eqref{ppij}. Other terms that must be taken into account, because of the same order of magnitude in the magnetic monopole charge, are the coupling between the magnetostatic field of the proton and the magnetic dipole momentum of the electron. The latter was already accounted in Ref.~\cite{M}. But differently from Ref.~\cite{M}, and consequently from Ref.~\cite{Bojowald:2018qqa}, the total Hamiltonian must necessarily encodes the magnetostatic interaction between the electron and the proton, and the interaction between the magnetostatic field generated by the proton and the magnetic momentum of the electron. In the rest of this letter, we take into account these two effects that cannot be neglected at the most relevant orders in $g$.\\

In natural units, the Hamiltonian of the electron in the Hydrogen atom that we must consider acquires the expression 
 \begin{eqnarray}
 \label{HH}
 H\, &&\, =\, T + V_{e} + V_{p} -\vec{\mu}_e \cdot \vec{B}_{p}  \nonumber \\
 &&=-\frac{1}{2 \mu}(\vec{\nabla}+ \imath e \vec{A})^{2}-\frac{e^{2}}{r}-\frac{g^2}{r} - \frac{e}{2 m_e}\, \frac{g}{r^2}\, \sigma_r,
 \end{eqnarray}
where $e$ and $g$ are (respectively) the electric and magnetic monopole charges in absolute value, $\vec{A}$ is the vector potential of the magnetic monopole, $\mu^{-1}=m_{e}^{-1} + M^{-1}$ is the reduced mass of the system ($m_{e}$ denoting the mass of the electron and $M$ the mass of the proton), $\vec{\mu}_e$ is the magnetic momentum of the electron and $\vec{B}_p$ is the magnetic field sourced by the magnetic monopole charge of the proton. The kinetic term is dictated by the non-associative momenta entering Eq.~\eqref{ppij}. 

\vspace{2mm}

It is common knowledge that the standard Schr\"odinger equation for the Hydrogen atom admits a wave function $\psi(r,\theta,\phi)$, which can be separated into two parts: one being a radial function $R_{nl}(r)$, and the other being an angular function, described by the spherical harmonics $Y_{lm}(\theta,\phi)$. Therefore, in absence of magnetic monopole charges, the total wave function reads $\psi(r,\theta,\phi)=R_{nl}(r)Y_{lm}(\theta,\phi)$. In the case under scrutiny here, we shall consider that the magnetostatic potential of the monopole charge has exactly the same form as the standard electrostatic potential, and thus conclude that the radial equation has the very same form as in the standard case. The structure of the angular-dependent components of the eigen-equation reads 
\begin{displaymath}
\mathcal{L}^2 =
\left( \begin{array}{cc}
L^2 +\gamma \cos\theta  & \gamma \sin\theta \,e^{\imath \phi}      \\
\gamma \sin\theta \,e^{\imath \phi}  &    L^2 -\gamma \cos\theta
\end{array} \right)
\end{displaymath}
where $\gamma=(\mu/m_e)\, e g \simeq eg$.
The matrix dependence of the Schr\"odinger equation was considered in Ref.~\cite{M}, where the eigenfunction were determined following the recipe provided in Ref.~\cite{F}. \\

{\it Case ii.a).--} We first focus on the case where the interaction between the magnetostatic field generated by the proton and the magnetic momentum of the electron is neglected. This amounts to formally set $\gamma=0$ in the previous relations. Consequently, the angular components of the eigenfunction still split as in the standard case $Y_{lm}(\theta,\phi)=\Theta_{lm}(\theta) e^{\imath m \phi}$, but with $\Theta_{lm}(\theta)$ to be opportunely determined, taking into account the deformed brackets among momenta and the magnetostatic interaction between the electron and the proton. Changing the angular variable into $x=(1-\cos\theta)/2$, and looking for a solution $\Theta(\theta)$ of the angular component of the new Shr\"odinger equation that has the form
\begin{equation}
\Theta(x)=x^{\frac{1}{2}|m|}(1-x)^{\frac{1}{2}|m-2eg|} u(x)\,,
\end{equation}
the angular Laplacian operator implies that $u(\theta)=\sum_n C_{n+s} \, x^{n+s}$, with $s=0, -|m|$ and 
\begin{eqnarray}
&& \frac{C_{n+1}}{C_n} = \frac{[n(n-1) + 2 (1+P)n +P (P+1) - (eg)^2 -\beta_0 ]}{[n(n+1) + (1+|m|) (n+1)]}  \nonumber   \,,\\
&& P=\frac{1}{2} \left( |m| + |m-2 eg | \right)\,.  \nonumber  
\end{eqnarray}
In the latter relations, 
\begin{eqnarray} 
\beta_0=l'(l'+1) - (eg)^2 \nonumber 
\end{eqnarray}
is the eigenstate of the operator $L^2$, which is expressed by 
\begin{eqnarray}
L^2\!=\!\frac{1}{\sin^2\theta} \left[ \sin \theta \, \frac{\partial}{\partial \theta} \, \sin \theta \, \frac{\partial}{\partial \theta} + \left\{  \frac{\partial}{\partial \phi} -\imath e g (1-\cos \theta)  \right\}^2  \right]. \nonumber 
\end{eqnarray}
In other words, as previously shown in Refs.~\cite{T,F}, the angular components eigen-equation casts 
$$L^2  \Theta(\theta) e^{\imath m \phi}= -\beta_0 \Theta(\theta) e^{\imath m \phi}\,,$$ 
provided that 
\begin{eqnarray} 
l'=n+P= |eg|\,, \ \  |eg| +1, \ \  |eg| +2 \ \ \cdots  \,. \nonumber 
\end{eqnarray}
The radial dependence is straightforwardly recovered from the Schr\"odinger equation 
\begin{equation}
\label{Eq}
\Big[-\frac{1}{2\mu r^{2}}\frac{d}{dr}\Big(r^{2}\frac{d}{dr}\Big)+\frac{L^{2}}{2\mu r^{2}}-\frac{e^{2}}{r}-\frac{g^2}{r}\Big]R_{nl}(r)=E\,R_{nl}(r)\, , \nonumber 
\end{equation}
where $L^{2}$ has eigenvalues 
$$l(l+1)= l'(l'+1)- (eg)^2\,.$$
Notice that the $g$-dependence entering the angular Laplacian relates to considering neither the electron-positron magnetostatic interaction, nor the interaction among the magnetic momentum of the electron and the magnetostatic field generated by the magnetic monopole of the proton. Instead, this is a feature of quantum mechanics in presence of a magnetic field. According to Eq.~(22) of Ref.~\cite{Bojowald:2018qqa}, this can be recovered by a simple shift in the angular Laplacian of the theory.

\vspace{2mm}

We notice that already at this level, solving for the radial equation would entail novel results than the ones hitherto discussed in the literature. An extra $g^2$-dependence is indeed easily achieved in the energy levels of the system. In natural units, the latter reads 
\begin{equation}
\label{Enp}
E_{n'}=-\frac{\mu}{2{n'}^{2}}(e^2+g^2)^2\,,
\end{equation}
hence retaining a non-vanishing first order expansion in $g^2$.
\\

{\it Case ii.b).--} We may now focus on the more general case in which the interaction between the magnetostatic field of the proton and the magnetic momentum of the electron is also considered. We first notice that $\mathcal{L}^2$ commutes with $\mathcal{J}_z= - \partial/\partial \phi + \sigma_z/2$. Because of the non diagonal form of $\mathcal{L}^2$, one may solve the equation 
\begin{equation}
\mathcal{L}^2 \Theta \Phi = -\beta \Theta \Phi \,.
\end{equation}
assuming
\begin{displaymath}
\Phi =
\left( \begin{array}{c}
e^{\imath (m-1) \phi}      \\
e^{\imath m \phi}    
\end{array} \right)\qquad  {\rm and}\qquad 
\Theta =
\left( \begin{array}{c}
\Theta_1 \\
\Theta_2
\end{array} \right) \,,
\end{displaymath}
with $m=0,\, \pm 1\, \pm 2\,, \cdots$. The eigen-equations for the angular Laplacian then cast 
\begin{eqnarray}
\Big[ && \sin \theta \, \frac{\partial}{\partial \theta} \, \sin \theta \, \frac{\partial}{\partial \theta} -\frac{1}{\sin \theta^2} \left\{ (m-1) - e g (1-\cos \theta)  \right\}^2 \nonumber \\
&&+\gamma \cos \theta  \Big]\Theta_1 + \gamma \sin \theta \Theta_2= -\beta \Theta_1\,, \\
\Big[ && \sin \theta \, \frac{\partial}{\partial \theta} \, \sin \theta \, \frac{\partial}{\partial \theta} -\frac{1}{\sin \theta^2} \left\{ m - e g (1-\cos \theta)  \right\}^2 \nonumber \\
&&-\gamma \cos \theta  \Big]\Theta_2 + \gamma \sin \theta \Theta_1= -\beta \Theta_2\,.
\end{eqnarray}
The system is then simplified making the {\it ans\"atze}
\begin{eqnarray}
\Theta_1(x)&=& x^{\frac{1}{2}|m|}(1-x)^{\frac{1}{2}|m-1-2eg|} u(x)\,, \nonumber \\
\Theta_2(x)&=& x^{\frac{1}{2}|m|}(1-x)^{\frac{1}{2}|m-2eg|} v(x)\,.  \nonumber
\end{eqnarray}
These finally allow us to recast the angular Laplacian eigen-equations, find the descending series in $x$ for $|u|$ and $|v|$
\begin{eqnarray}
|u|=x^n+ C_1 x^{n-1}+ \cdots \,, |v|=x^n+ C_2 x^{n-1}+ \cdots \,,
\end{eqnarray}
with $C_1$, $C_2$ generic coefficients that are eliminated while recovering the general form of the eigenvalues 
\begin{eqnarray}\label{aut}
\beta= {l''}^2 - (eg)^2 \pm \left[ {l''}^2 - (eg)^2 + (eg- ^2\right]^{\frac{1}{2}}
\,,
\end{eqnarray}
where $l''=n+P+\epsilon$, $P$ having been defined above and with $\epsilon=1$ for $(m-2 (eg))\leq 0$ and $m\leq 0$, otherwise zero.  
From Eq.~\eqref{aut} it is straightforward to distinguish that the lowest roots compatible with the case in which $\gamma$ is formally set to zero are 
\begin{eqnarray}
\beta= |eg| - \frac{eg}{|eg|} \gamma\,.
\end{eqnarray}

The radial dependence can be easily solved accounting for the usual definitions, extended in order to encode the magnetostatic central interaction, namely $k^2= -2\mu/E$ and $U(r)= - 2 \mu (e^2+g^2)/r$, which leads to the radial part of the Shr\"odeinger equation
\begin{eqnarray}
\frac{1}{r^2} \frac{\partial}{\partial r} r^2 \frac{\partial}{\partial r} R -\left(k^2 +U(r) +\frac{\beta}{r^2}\right) R=0
\,.
\end{eqnarray}
Changing radial variable into $\rho= 2 k r$, and letting $n'=\mu (e^2+g^2)/k$, as for the Laguerre polynomials, one finds  
\begin{eqnarray}
\frac{1}{\rho^2} \frac{\partial}{\partial \rho} \rho^2 \frac{\partial}{\partial \rho} R + \left(-\frac{1}{4} +\frac{n'}{\rho} -\frac{\beta}{\rho^2}\right) R=0
\,,
\end{eqnarray}
which asymptotically, for $R= e^{-\frac{1}{2} \rho} F$, provides  
\begin{eqnarray}
\frac{\partial^2 \! F}{\partial \rho^2}+\left( \frac{2}{\rho} -1 \right) \frac{\partial \! F}{\partial \rho} + \left[\frac{n'-1}{\rho} - \frac{\beta}{\rho^2} \right] F=0\,.
\,,
\end{eqnarray}
Recasting $F(\rho)=\rho^s L(\rho)$, with 
\begin{eqnarray}
s=\frac{-1 +(1+4\beta)^{\frac{1}{2}}}{2}= \left| |eg| \pm [l(l+1)]^{\frac{1}{2}} \right| - |eg| \,.
\end{eqnarray}
This finally leads us to derive the equation 
\begin{eqnarray}
\rho \frac{\partial^2 \! L}{\partial \rho^2}+ \left[ 2(s+1) -\rho \right] \frac{\partial  L}{\partial \rho} + (n' -s -1) L=0\,,
\end{eqnarray}
which implies that $L$ is a polynomial of order $n''$, if 
\begin{equation}
n'=n''+s+1=n'' + |eg| + \left| |eg| \pm [l(l+1)]^{\frac{1}{2}} \right|\,.
\end{equation}
Consequently, the radial eigen-function is recovered $R=e^{-\frac{1}{2}\rho} \rho^s L(\rho)$, and the eigenvalues as Eq.(\ref{Enp}).

\vspace{3mm}

We emphasize again that the magnetostatic interaction potential is absent in the analysis carried out in Ref.~\cite{Bojowald:2018qqa}, since it cannot be incorporated in the centrifugal barrier term $L^{2}/(2 \mu r^{2})$, as a redefinition of the angular momentum operators 
\begin{equation}
\label{LpL}
\vect{L}'=\vect{L}+eg \vect{r}/r\, .
\end{equation}
These latter retain the same algebra as the $\vect{L}$ operators, but are also related to the standard Casimir operator $L^{2}$, merely by the constant shift
\begin{equation}
\label{LLppp}
L^{2}={L'}^{2}-e^{2}g^{2}\,,
\end{equation}
which corresponds to Eq.~(22) of Ref.~\cite{Bojowald:2018qqa}. As argued by Bojowald {\it et al.}, the non-associativity induced by the presence of magnetic monopole charges totally amounts to this shift in the angular components of the Laplacian operator, {\it i.e.} 
\begin{equation}
\label{shift}
\frac{L'^{2}}{2\mu r^{2}}=\frac{L^{2}}{2 \mu r^{2}}+\frac{e^2g^2}{2 \mu r^{2}}\, . 
\end{equation}
The last term in Eq.~\eqref{shift} has a different dependence on the radius than the magnetostatic interaction potential, 
and also a different sign. \\

We also emphasize that the energy levels of the Hydrogen atom are only labelled by the $n'$ quantum numbers, and in any circumstance by the quantum numbers $l$ and $m$. For the standard case, it is very well known that the spectrum of the Hydrogen atom only retains a $n$-eigenvalues dependence, associated with the electrostatic potential. It is also relevant that redefinition of the $n$-eigenvalues that encode $g$-dependence are not observable (we may hence neglect the prime in the definition of the $n$ quantum numbers) and hence cannot be used to pose constraints on the magnetic monopole charge of the proton and/or of the electron. Finally, by virtue of the symmetric situation among the electrostatic and the magnetostatic fields, we have shown here what was expected, {\it i.e.} that the central magnetostatic potential added to the Schr\"odinger equation again only affects the $n$-dependence of the energy levels.   

\vspace{1mm}

{\it Discussion.--} We derived new stronger bounds on the magnetic monopole charge $g$ of dyonic protons, and consequently on non-associative quantum mechanics.
The strongest limit is obtained from measures of the average magnetic field of the Moon: 
\begin{eqnarray}
g \leq 7.38 \times 10^{-36}g_{\rm Dirac}\,.
\end{eqnarray}
\vspace{1mm}

The new limits are so strong to constrain an eventual magnetostatic force to be weaker than gravity. This can be of interest in light of the celebrated {\it Weak Gravity Conjecture} --- see {\it e.g.} Ref.~\cite{ArkaniHamed:2006dz}. If we assume that electrons are also dyons, we further conclude that the monopole magnetostatic field would affect the energy levels of the Hydrogen atom, not by virtue of the $l$ and $m$ numbers' dependence on the atomic levels, which instead remains unchanged, but of the $n$ quantum numbers. We manage to estimate the correct bound on $g$, pondering the following facts: i) the energy levels of the new system are $E_{n}=-(\mu/2n^{2})(e^2+g^2)^2$; ii) the limits on the Hydrogen atom spectroscopy entail $\Delta E_{1s-2s}/E_{1s-2s}\simeq 4.5 \times 10^{-15}$ \cite{12}. Consequently, these considerations percolate into a lower experimental constraint on $g$ than the tightest one estimated in Ref.~\cite{Bojowald:2018qqa} (by approximately one order of magnitude), namely  
\begin{equation}
\label{g}
g  \leq 1.5 \times 10^{-8}   g_{\rm Dirac}\,,
\end{equation} 
which is expressed in terms of the smallest Dirac magnetic charge.

\vspace{-2mm}

\acknowledgements 
\noindent
We acknowledge private conversations with M. Bojowald on this subject. 
We wish to acknowledge support by the NSFC, through the grant No. 11875113, the Shanghai Municipality, through the grant No. KBH1512299, and by Fudan University, through the grant No. JJH1512105.

\end{document}